# Influence of Electron-electron Drag on Piezoresistance of *n*-Si

## I.I. Boiko[1]

*Institute of Semiconductor Physics, NAS of Ukraine, 45 prospect Nauky, 03028 Kyiv, Ukraine*
(Dated: February 1, 2011)

Piezoresistance of *n*-Si is considered with due regard for inter-valley drag. It is shown that intervalley drag gains the piezocoefficient and diminishes the mobility. In the region of nondegenerate carriers the effect of drag increases when carrier concentration rises and temperature falls.

## 1. Introduction

. In crystals with one and only simple band the electron-electron scattering does not change total momentum of carriers and therefore does not give a direct, independent contribution in the conductivity. Quite other situation we have for crystals with a composite band structure. There the conductivity of crystal can be essentially influenced by mutual drag of carriers, which belong to different partial bands or valleys (see Refs. [1, 2]). In particular, the inter-valley drag can sufficiently diminish electron mobility of *n*-Si and *n*-Ge at low temperatures. The reason of that is principal difference between scattering of band electrons from some valley on fixed charged impurities or equilibrium phonons and scattering on nonequilibrium electrons from other valley, where divergence of equilibrium is distinct one. Now we have some right to hope that in multy-valley semiconductors the inter-valley drag can noticeably influence not only on conductivity but on piezoresistance as well. In this article we will pay again the main attention to region of low temperatures where coulomb scattering is not damped by collisions of electrons with phonons. We restrict here our calculations by charged impurities and acoustic phonons as external scattering system.

## 2. Balance equations

In Refs [1, 2] it was presented the set of balance equations obtained as a first momentum of quantum kinetic equations:

$$e \vec{E} + \vec{F}^{(a)} + \sum_{b=1}^{6} \vec{F}^{(a,b)} = 0, \qquad (a = 1, 2, \ldots , 6) \qquad (1)$$

Here symbol *a* numerates different valleys; $\vec{E}$ is applied electric field; the resistant force

$$\vec{F}^{(a)} = -\frac{e^2}{(2\pi)^6 n_a} \int d^3\vec{k} \int \vec{q} d^3\vec{q} \int d\omega \, \delta(\hbar\omega - \varepsilon_{\vec{k}}^{(a)} + \varepsilon_{\vec{k}-\vec{q}}^{(a)}) \{f_{\vec{k}}^{(a)} - f_{\vec{k}-\vec{q}}^{(a)} + [f_{\vec{k}}^{(a)}(1 - f_{\vec{k}-\vec{q}}^{(a)}) +$$

$$+ f_{\vec{k}-\vec{q}}^{(a)}(1 - f_{\vec{k}}^{(a)})]\tanh(\hbar\omega/2k_B T)\}[\langle \varphi^{2}_{(I)} \rangle_{\omega,\vec{q}} + \langle \varphi^{2}_{(ph)} \rangle_{\omega,\vec{q}}] \qquad (2)$$

---

[1] E-mail: igorboiko@yandex.ru

is related to interaction of band carriers with charged impurities disposed uniformly in space; $\langle \varphi^2_{(I)} \rangle_{\omega,\vec{q}}$ and $\langle \varphi^2_{(ph)} \rangle_{\omega,\vec{q}}$ are Fourier components of correlator of impurity and phonon scattering potentials. The drag force

$$\vec{F}^{(a,b)} = \frac{e^4 \hbar}{4\pi^6 n^{(a)}} \int \vec{k}\, d^3\vec{k} \int d^3\vec{k}' \int d^3\vec{q}\, \frac{1}{q^4} \frac{\delta(\varepsilon^{(a)}_{\vec{k}} - \varepsilon^{(a)}_{\vec{k}-\vec{q}} - \varepsilon^{(b)}_{\vec{k}'} + \varepsilon^{(b)}_{\vec{k}'-\vec{q}})}{|\varepsilon(\omega=0,\vec{q})|^2} Y_{ab}(\vec{k},\vec{k}',\vec{q})\ ;$$

$$Y_{ab}(\vec{k},\vec{k}',\vec{q}) = f^{(a)}_{\vec{k}-\vec{q}}(1-f^{(a)}_{\vec{k}}) f^{(b)}_{\vec{k}'}(1-f^{(b)}_{\vec{k}'-\vec{q}}) - f^{(a)}_{\vec{k}}(1-f^{(a)}_{\vec{k}-\vec{q}}) f^{(b)}_{\vec{k}'-\vec{q}}(1-f^{(b)}_{\vec{k}'}) \quad (3)$$

relates to interaction between drifting carriers from $a$- and b-valleys.

Here $n^{(a)}$ and $\varepsilon^{(a)}_{\vec{k}}$ are concentration and dispersion law for electrons from $a$-valley. For undeformed crystal of $n$-Si we have (see Fig. 1):

$$\varepsilon^{(a)}_{\vec{k}} = \frac{\hbar^2}{2}\left(\frac{k_x^2}{m^{(a)}_{xx}} + \frac{k_y^2}{m^{(a)}_{yy}} + \frac{k_z^2}{m^{(a)}_{zz}}\right). \quad (4)$$

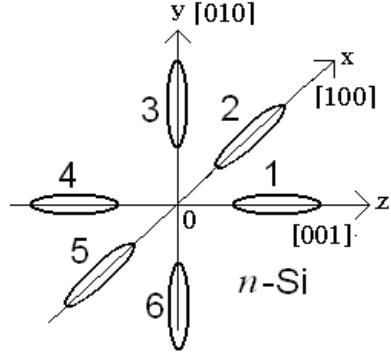

Fig. 1. Band structure of $n$-silicon.

For two valleys ($a$ = 1 and 4) $m_{zz} = m_\parallel$, $m_{xx} = m_{yy} = m_\perp$; for two valleys ($a$ = 2 and 5) $m_{xx} = m_\parallel$, $m_{zz} = m_{yy} = m_\perp$ and for two valleys ($a$ = 3 and 6) $m_{yy} = m_\parallel$, $m_{xx} = m_{zz} = m_\perp$. Therefore

$$\varepsilon^{(1,4)}_{\vec{k}} = \frac{\hbar^2}{2}\left(\frac{k_x^2+k_y^2}{m_\perp} + \frac{k_z^2}{m_\parallel}\right);\ \varepsilon^{(2,5)}_{\vec{k}} = \frac{\hbar^2}{2}\left(\frac{k_z^2+k_y^2}{m_\perp} + \frac{k_x^2}{m_\parallel}\right);\ \varepsilon^{(3,6)}_{\vec{k}} = \frac{\hbar^2}{2}\left(\frac{k_x^2+k_z^2}{m_\perp} + \frac{k_y^2}{m_\parallel}\right). \quad (5)$$

The screening dielectric function for quasi-elastic collisions has the form

$$\varepsilon(\omega,\vec{q}) = \varepsilon_L + \Delta\varepsilon_e(\omega=0,\vec{q}),$$

where $\varepsilon_L$ is dielectric constant of crystal lattice and $\Delta\varepsilon_e(\omega,\vec{q})$ is contribution of band electrons in total dielectric function. For convenience we will use the following approximate form:

$$\Delta\varepsilon_e(0,\vec{q}) = \varepsilon_L q_0^2(\vec{q})/q^2 .$$

Then

$$\langle \varphi^2_{(I)} \rangle_{\vec{q},\omega} = \frac{32\pi^3 e^2 n_I}{\varepsilon_L^2 [q^2 + q_0^2(\vec{q})]^2}\delta(\omega), \quad \langle \delta\varphi^2_{(ph)} \rangle_{\vec{q},\omega} = \Xi_A^2 \frac{2\pi k_B T}{e^2 \rho s^2}\delta(\omega) . \tag{6}$$

Here $n_I$ is concentration of charged impurities, $\Xi_A = \Xi_d + (1/2)\Xi_u$ where $\Xi_d$ and $\Xi_u$ are dilation and shear deformation potential constants (see, for example, Refs. [6] and [7]). The form (7) corresponds to the approximation of quasielastic collisions.

The screening plays a significant, even appointing role in the area of small transferred vectors $q$; therefore instead of $q_0^2(\vec{q})$ we can use the following expression (see Ref. [1]):

$$q_0^2(q) \to q_0^2(0) = \frac{12 e^2 m_{\parallel}^{3/2} \sqrt{2k_B T}}{\hbar^3 L\sqrt{\pi}\varepsilon_L} F_{-1/2}(\eta) . \tag{7}$$

Here Fermi-integral

$$F_r(\eta) = \frac{1}{\Gamma(r+1)}\int_0^\infty \frac{w^r\, dw}{1+\exp(w-\eta)} , \tag{8}$$

$L = m_{\parallel}/m_{\perp}$, $\Gamma$ is gamma-function, $\eta = \varepsilon_F / k_B T$ is dimensionless Fermi-energy. The form (7) is valid for deformed crystal if one uses linear approximation over deformation. For nondegenerate carriers

$$q_0^2(0) = 4\pi e^2 n / \varepsilon_L k_B T . \tag{9}$$

To calculate drift velocities $\vec{u}^{(a)}$ of electrons from $a$-group we accept the model of non-equilibrium distribution functions as Fermi functions with argument containing shift of velocity $\vec{v}^{(a)}(\vec{k}) = \hbar^{-1}(\partial\varepsilon_{\vec{k}}^{(a)}/\partial\vec{k})$ on correspondent velocity $\vec{u}^{(a)}$:

$$f_{\vec{k}}^{(a)} = f^{0(a)}(\vec{v}^{(a)}(\vec{k}) - \vec{u}^{(a)}). \qquad (a = 1, 2, \ldots, 6) \tag{10}$$

Here $f^{0(a)}(\vec{v}^{(a)}(\vec{k})) = f_0^{(a)}(\varepsilon)$ is equilibrium distribution function for $a$-carriers. Drift velocities $\vec{u}^{(a)}$ are proportional to partial densities of currents $\vec{j}^{(a)}$:

$$\vec{u}^{(a)} = \frac{1}{en^{(a)}}\vec{j}^{(a)} . \tag{11}$$

Here $n^{(a)}$ is concentration of electrons in $a$-valley. The density of total current:

$$\vec{j} = \sum_{a=1}^{6}\vec{j}^{(a)} . \tag{12}$$

Using the forms (10) and carrying out linearization of forces in Eqs. (2 and 3) over drift velocities we obtain:

$$\vec{F}^{(a)} = -e\,\tilde{\beta}^{(a)}\,\vec{u}^{(a)}\;;\quad \vec{F}^{(a,b)} = -e\,\tilde{\xi}^{(a,b)}\left(\vec{u}^{(a)} - \vec{u}^{(b)}\right). \tag{13}$$

Here components of tensors $\tilde{\beta}^{(a)}$ and $\tilde{\xi}^{(a,b)}$ are (see [2, 3] and Eqs. (6) and (7)):

$$\beta_{uv}^{(a)} = \lambda_{uv}^{(a)} + \chi_{uv}^{(a)}\;; \tag{14}$$

$$\lambda_{uv}^{(a)} = \frac{\hbar}{2(2\pi)^5\,e\,n^{(a)}\,k_B T}\int d\omega \int \frac{q^2\,d^3\vec{q}}{\sinh(\hbar\omega/k_B T)} q_u q_v \,\mathrm{Im}\Delta\varepsilon^{0}_{(a)}(\omega,\vec{q})\langle\varphi^2_{(I)}\rangle_{\omega,\vec{q}} =$$

$$= \frac{en_I}{2\pi^2\,n^{(a)}\varepsilon_L^2}\int \left[\frac{1}{\omega}\mathrm{Im}\Delta\varepsilon^{0}_{(a)}(\omega,\vec{q})\right]_{\omega=0} \frac{q^2 q_u q_v}{[q^2 + q_0^2(\vec{q})]^2} d^3\vec{q}\;; \tag{15}$$

$$\chi_{uv}^{(a)} = \frac{\hbar}{2(2\pi)^5\,e\,n^{(a)}\,k_B T}\int d\omega \int \frac{q^2\,d^3\vec{q}}{\sinh(\hbar\omega/k_B T)} q_u q_v \,\mathrm{Im}\Delta\varepsilon^{0}_{(a)}(\omega,\vec{q})\langle\varphi^2_{(ph)}\rangle_{\omega,\vec{q}} =$$

$$= \frac{\Xi_A^2\,k_B T}{32\pi^4 n^{(a)} e^3 \rho s^2}\int \left[\frac{1}{\omega}\mathrm{Im}\Delta\varepsilon^{0}_{(a)}(\omega,\vec{q})\right]_{\omega=0} q^2 q_u q_v\,d^3\vec{q}\;; \tag{16}$$

$$\xi_{uv}^{(a,b)} = \frac{\hbar^2(1-\delta_{ab})}{2(2\pi)^4\,e\,\varepsilon_L^2 n^{(a)} k_B T}\int \frac{d\omega}{\sinh^2(\hbar\omega/2k_B T)}\int \frac{q^4 q_u q_v d^3\vec{q}}{[q^2 + q_0^2(q)]^2}\mathrm{Im}\Delta\varepsilon^{0}_{(a)}(\omega,\vec{q})\mathrm{Im}\Delta\varepsilon^{0}_{(b)}(\omega,\vec{q})\;. \tag{17}$$

In Eqs. (15) – (17) the indices $u,v = x,y,z$ and the imaginary part of dielectric function

$$\mathrm{Im}\Delta\varepsilon^{0}_{(a)}(\omega,\vec{q}) = -\frac{e^2}{\pi q^2}\int d^3\vec{k}[f_0(\varepsilon_{\vec{k}}^{(a)}) - f_0(\varepsilon_{\vec{k}-\vec{q}}^{(a)})]\,\delta(\varepsilon_{\vec{k}-\vec{q}}^{(a)} - \varepsilon_{\vec{k}}^{(a)} + \hbar\omega)\;.$$

For quasielastic collisions we have the form (see [1])

$$\mathrm{Im}\Delta\varepsilon^{(0)}_{(a)}(\omega\to 0,\vec{q}) = \omega\,\Lambda^{(a)}(\vec{q})\;,$$

where

$$\Lambda^{(a)}(\vec{q}) = \frac{2e^2\,m_\perp m_\parallel^{1/2}}{q^2\hbar^3}\left(\frac{q_x^2}{m_{xx}^{(a)}} + \frac{q_y^2}{m_{yy}^{(a)}} + \frac{q_z^2}{m_{zz}^{(a)}}\right)^{-1/2}\left[1 + \exp\left\{\frac{\hbar^2}{8k_B T}\left(\frac{q_x^2}{m_{xx}^{(a)}} + \frac{q_y^2}{m_{yy}^{(a)}} + \frac{q_z^2}{m_{zz}^{(a)}}\right) - \eta_a\right\}\right]^{-1}. \tag{18}$$

As result we have:

$$\lambda_{uv}^{(a)} = \frac{en_I}{2\pi^2\,n^{(a)}\varepsilon_L^2}\int \Lambda^{(a)}(\vec{q})\frac{q^2 q_u q_v}{[q^2 + q_0^2(\vec{q})]^2}d^3\vec{q}\;; \tag{19}$$

$$\chi_{uv}^{(a)} = \frac{\Xi_A^2 k_B T}{32n^{(a)}\pi^2 e^3 \rho s^2}\int \Lambda^{(a)}(\vec{q})q^2 q_u q_v d^3\vec{q}\;; \tag{20}$$

$$\xi_{uv}^{(a,b)} = \frac{\gamma(k_B T)^2}{4\hbar\pi^4 e\,\varepsilon_L^2 n^{(a)}}\int \frac{q^4 q_u q_v}{[q^2 + q_0^2(q)]^2}\Lambda^{(a)}(\vec{q})\Lambda^{(b)}(\vec{q})d^3\vec{q}, \tag{21}$$

where

$$\gamma = \int_{-\infty}^{\infty}\frac{w^2\,dw}{\sinh^2 w}\approx 3.29\;. \tag{22}$$

Note, that

$$n^{(a)} \xi^{(a,b)}_{\alpha\beta} = n^{(b)} \xi^{(b,a)}_{\alpha\beta} . \quad (23)$$

Farther we assume $n_I = n = \sum_{a=1}^{6} n^{(a)}$.

From Eqs. (1) and (13) one obtains the system of equations for partial drift velocities:

$$\vec{E} - \tilde{\beta}^{(a)} \vec{u}^{(a)} - \sum_{b=1}^{6} \tilde{\xi}^{(a,b)} \left( \vec{u}^{(a)} - \vec{u}^{(b)} \right) = 0 . \quad (a = 1, 2, \ldots, 6) \quad (24)$$

In these system of equations kinetic coefficients $\beta^{(a)}$ and $\xi^{(a,b)}$ have a matrix form, matrices $\tilde{\xi}^{(a,b)}$ responding for inter-valley drag. The value $(\tilde{\beta}^{(a)})^{-1}$ is mobility tensor for $a$-carriers if one neglects inter-valley drag ($\tilde{\xi}^{(a,b)} \to 0$).

## 3. Populations of valleys in deformed crystal

Let a silicon crystal is mechanically compressed along axis [001] (see Fig. 1), then components of the stress tensor are

$$X_{\alpha\beta} = -\delta_{\alpha\beta} \delta_{\alpha z} X \text{ (here } X > 0 \text{ )}. \quad (25)$$

For this situation dispersion law for different valleys has the following form (see Ref. [4]):

$$\varepsilon^{(a)}_{\vec{k}}(X) = \varepsilon^{(a)}_{\vec{k}}(X=0) + \Delta\varepsilon^{(a)}(X) \equiv \varepsilon^{(a)}_{\vec{k}} + \Delta\varepsilon^{(a)}(X) , \quad (a=1, 2, \ldots, 6) \quad (26)$$

where

$$\Delta\varepsilon^{(1)}(X) = \Delta\varepsilon^{(4)}(X) = -\frac{2}{3} \Xi_u (s_{11} - s_{12}) X ,$$

$$\Delta\varepsilon^{(2)}(X) = \Delta\varepsilon^{(5)}(X) = \Delta\varepsilon^{(3)}(X) = \Delta\varepsilon^{(6)}(X) = \frac{1}{3} \Xi_u (s_{11} - s_{12}) X , \quad (27)$$

$\Xi_u$ is the shear deformation potential, $s_{11}$ and $s_{12}$ are the elastic constants.

Density of carriers in $a$-valley is

$$n^{(a)}(X) = \int \frac{d^3\vec{k}}{4\pi^3} f_0^{(a)}(\varepsilon^{(a)}_{\vec{k}}(X)) = \int \frac{d^3\vec{k}/4\pi^3}{1+\exp[\varepsilon^{(a)}_{\vec{k}}(X)/k_B T - \eta]} = \int \frac{d^3\vec{k}/4\pi^3}{1+\exp[(\varepsilon_{\vec{k}} + \Delta\varepsilon^{(a)}_{\vec{k}}(X))/k_B T - \eta]} . \quad (28)$$

For nondegenerate carriers

$$n^{(a)}(X) = (n/6)[1 - \Delta\varepsilon^{(a)}(X)/k_B T] . \quad (29)$$

It follows from conservation of total number of carriers (see Eq. (8)):

$$\sum_{a=1}^{6} n^{(a)}(X) = \sum_{a=1}^{6} n^{(a)}(0) = n = \frac{3(2\pi k_B T)^{3/2} m_{\parallel}^{3/2}}{2\pi^3 \hbar^3 L} F_{1/2}(\eta) . \quad (30)$$

## 4. Kinetic coefficients for deformed silicon crystal

Consider the case when applied electric field is applied along some fourfold *l*-axis, that is

$$\vec{E} = E\,\vec{e}_l. \qquad (l = x, y, z) \qquad (31)$$

Then only *l*-components of drift velocities $\vec{u}^{(a)}$ and diagonal components of tensors $\tilde{\beta}^{(a)}$ and $\tilde{\xi}^{(a,b)}$ are distinct of zero (in coordinate system related to fourfold axes) and then one can write the system of equations corresponding to the system. (24) in the form

$$E_l - \beta_{ll}^{(a)}(X)u_l^{(a)} - \sum_{b=1}^{6}\xi_{ll}^{(a,b)}(X)\left(u_l^{(a)} - u_l^{(b)}\right) = 0 \ . \ (a = 1, 2, \ldots, 6) \quad (32)$$

Here the expressions for $\beta_{ll}^{(a)}(X)$ and $\xi_{ll}^{(a,b)}(X)$ have the forms (14), (18) – (21) where dimensionless Ferny energy $\eta_a$ in formulae (18) is shifted by such way;

$$\eta_a \to \eta - \Delta\varepsilon^{(a)}(X)/k_B T \ . \qquad (33)$$

Farther we assume the deformation to be small and linearize all expressions over stress $X$.

In this article we consider only nondegenerate gas. Then it follows from the symmetry of considered system of carriers:

$$n^{(a)}(X) = (n/6)\,C_a(X);\quad C_4(X) = C_1(X);\ C_3(X) = C_5(X) = C_6(X) = C_2(X); \quad (34)$$

where (see Eqs. (29), (30))

$$C_1(X) = 1 + 2\Xi_u(s_{11} - s_{12})X/3k_B T\,; \quad C_2(X) = 1 - \Xi_u(s_{11} - s_{12})X/3k_B T \ . \qquad (35)$$

Note, that $C_1(X) + 2C_2(X) = 3$. It follows also:

$$\beta_{ll}^{(a)}(X) = \beta_{ll}^{(a)}(X=0)\ ; \ \xi_{ll}^{(a,b)}(X) = \xi_{ll}^{(a,b)}(X=0)\,C_b(X). \qquad (36)$$

Consider now the case when applied electric field is applied along *z*-axis, that is

$$\vec{E} = E\,\vec{e}_z. \qquad (37)$$

Then only *z*-components of drift velocities $\vec{u}^{(a)}$ and *zz*-components of tensors $\tilde{\beta}^{(a)}$ and $\tilde{\xi}^{(a,b)}$ are distinct of zero and one can write Eqs. (32) in the form

$$E_z - \beta_a u_z^{(a)} - \sum_{b=1}^{6}\xi^{(a,b)}C^{(b)}(X)\left(u_z^{(a)} - u_z^{(b)}\right) = 0 \ . \qquad (a = 1, 2, \ldots, 6) \qquad (38)$$

Here $\beta_a = \beta_{zz}^{(a)}(X=0)\,; \xi^{(a,b)} = \xi_{zz}^{(a,b)}(X=0)$.

It is evident, that

$$u_z^{(1)} = u_z^{(4)}\,; \quad u_z^{(2)} = u_z^{(5)} = u_z^{(3)} = u_z^{(6)} \qquad (39)$$

and so on. Solving the system (38) and using the relations (39) we obtain ($\xi^{(1,2)} \to \xi$)

$$u_z^{(1,2)}(X;\xi) = E_z \frac{\beta_{2,1} + 6\xi}{\beta_1\beta_2 + 2\xi[\beta_1 C_1(X) + 2\beta_2 C_2(X)]} . \tag{40}$$

One can write in the following form the resulting expression for total conductivity of deformed crystal:

$$\sigma_{zz}(X;\xi) = \frac{j_z(X;\xi)}{E_z} = \frac{1}{E_z}\sum_{a=1}^{6} j_z^{(a)}(X;\xi) = \frac{e}{E_z}\sum_{a=1}^{6} n^{(a)}(X)\, u_z^{(a)}(X;\xi) = en\mu_{zz}(X;\xi) =$$

$$= \frac{en}{3}\frac{\beta_2 C_1(X) + 2\beta_1 C_2(X) + 18\,\xi}{\beta_1\beta_2 + 2\xi\,[\beta_1 C_1(X) + 2\beta_2 C_2(X)]} . \tag{41}$$

For undeformed crystal

$$\sigma_{zz}(0;\xi) = en\mu(\xi) = \frac{en}{3}\frac{\beta_2 + 2\beta_1 + 18\xi}{\beta_1\beta_2 + 2\xi\,(\beta_1 + 2\beta_2)} . \tag{42}$$

Determine the piezoresistance coefficient $\pi_{kk}(\xi)$ by the expression (see Refs. [5, 6])

$$\pi_{kk}(\xi) = -\frac{1}{\sigma_{kk}(0;\xi)}\left[\frac{d\sigma_{kk}(X;\xi)}{dX}\right]_{X=0} . \tag{43}$$

If we direct electrical field $\vec{E}$ along $x$- and $y$-axes we will obtain for the case (25) the relation

$$\pi_{zz}(\xi) = -2\pi_{xx}(\xi) = -2\pi_{yy}(\xi) . \tag{44}$$

To investigate the dependence of conductivity and piezoresistance coefficient on parameter of intervalley drag we use the following formulae obtained from the Eqs. (35), (42), (43):

$$\frac{\sigma_{zz}(0;\xi)}{\sigma_{zz}(0;0)} = \frac{\mu(\xi)}{\mu(0)} = \frac{\beta_1\beta_2(\beta_2 + 2\beta_1 + 18\xi)}{(\beta_2 + 2\beta_1)[\beta_1\beta_2 + 2\xi(\beta_1 + 2\beta_2)]} ; \tag{45}$$

$$\sigma_{zz}(0,0) = \frac{en}{3}\frac{\beta_2 + 2\beta_1}{\beta_1\beta_2} ; \tag{46}$$

$$\frac{\pi_{zz}(\xi)}{\pi_{zz}(0)} = \frac{(\beta_2 + 2\beta_1)[\beta_1\beta_2 + 6\xi(\beta_1 + \beta_2) + 36\xi^2]}{(\beta_2 + 2\beta_1 + 18\xi)[\beta_1\beta_2 + 2\xi(\beta_1 + 2\beta_2)]} ; \tag{47}$$

$$\pi_{zz}(0) = \frac{\Xi_u(s_{11} - s_{12})}{3k_B T}\frac{2(\beta_2 - \beta_1)}{\beta_2 + 2\beta_1} . \tag{48}$$

### 5. Results of numerical calculations

These results are shown on Figs. 2 – 4. Here Fig. 2(a, b, c) shows absolute value of piezoresistance coefficient. We use for calculations such data (see Refs. [5], [6]):

$m_0 = 9.1066 \cdot 10^{-28} g$, $m_\| = 8.342 \cdot 10^{-28} g$; $M = 6$; $L = 4.8$; $\varepsilon_L = 12$; $(m_\| / m_0) = 0.916$; $s_{11} - s_{12} = 9.82 \cdot 10^{-12} Pa^{-1}$, $\Xi_u = 8.6 eV$; $\varepsilon_L = 12$, $(m_\| / m_0) = 0.92$, $\rho s^2 = 1.66 \cdot 10^{11} Pa$, $\Xi_A = \Xi_d + (1/2)\Xi_u = -4.2 eV$. On figures (a), (b), (c) the solid lines represent piezoresistance coefficient for crystal where band carriers are involved in drag. The dashed lines correspond to the calculations which ignore the intervalley drag.

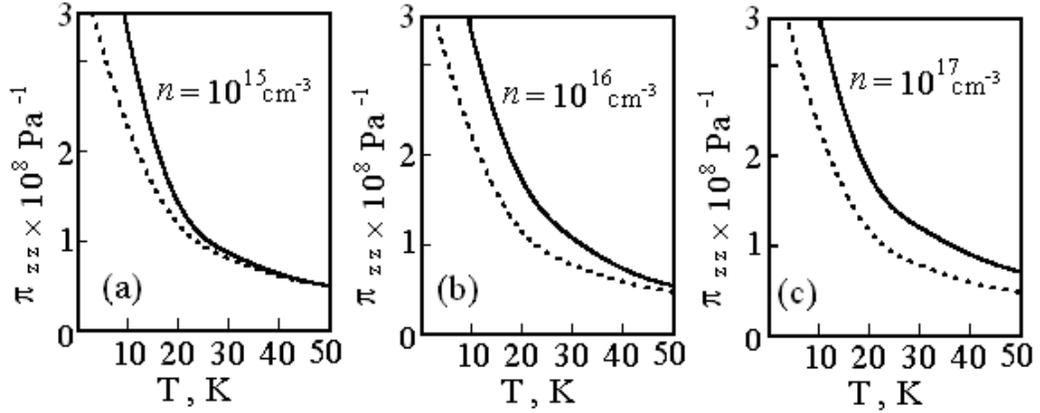

Fig. 2. Dependence of relative piezocefficient $\pi_{zz}$ on temperature.

Figs. 3 and 4 represent relative values. One can see that intervalley drag gains the piezocefficient and diminishes the mobility. In the region of nondegenerate carriers the effect of drag increases when carrier concentration rises and temperature falls.

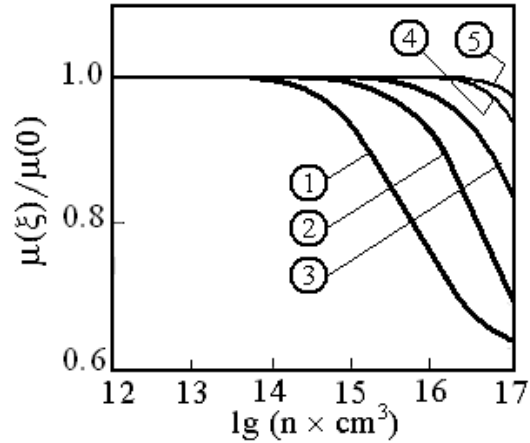

Fig. 3. Dependence of relative mobility on carrier concentration.

$1 - T = 20 K$; $2 - T = 40 K$; $3 - T = 70 K$; $4 - T = 120 K$; $5 - T = 150 K$.

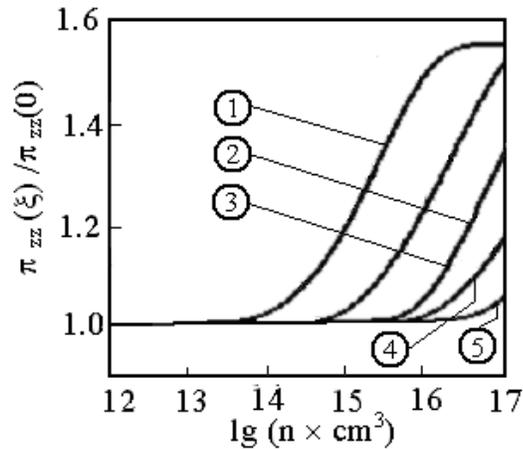

Fig. 4. Dependence of relative piezocefficient on carrier concentration.

$1 - T = 20\,K$; $2 - T = 40\,K$; $3 - T = 70\,K$; $4 - T = 100\,K$; $5 - T = 150\,K$.

---